\documentclass[epj,twocolumn]{webofc}
\usepackage[varg]{txfonts}   % Web of Conferences font
\usepackage{subfig}
\woctitle{The Innermost Regions of Relativistic Jets and Their Magnetic Fields}
\begin{document}
\title{ 	Imaging VLBI polarimetry data from Active Galactic Nuclei using the Maximum Entropy Method}

\author{Colm P. Coughlan\inst{1}\fnsep\thanks{\email{colm.coughlan@umail.ucc.ie}} \and
        Denise C. Gabuzda\inst{1}\fnsep\thanks{\email{d.gabuzda@ucc.ie}}
}

\institute{Department of Physics, University College Cork, Ireland}

\abstract{
Mapping the relativistic jets emanating from AGN requires the use of a deconvolution algorithm to account for the effects of missing baseline spacings. The CLEAN algorithm is the most commonly used algorithm in VLBI imaging today and is suitable for imaging polarisation data. The Maximum Entropy Method (MEM) is presented as an alternative with some advantages over the CLEAN algorithm, including better spatial resolution and a more rigorous and unbiased approach to deconvolution. We have developed a MEM code suitable for deconvolving VLBI polarisation data. Monte Carlo simulations investigating the performance of CLEAN and the MEM code on a variety of source types are being carried out. Real polarisation (VLBA) data taken at multiple wavelengths have also been deconvolved using MEM, and several of the resulting polarisation and Faraday rotation maps are presented and discussed.  
}
\maketitle

\section{The Maximum Entropy Method}
\label{MEM}

The Maximum Entropy Method (MEM) is an alternative deconvolution method to the CLEAN algorithm. It is a constrained optimisation method, and is based on a consideration of the function

\begin{equation}
\label{jeqn}
\begin{split}
J = H(I_{m},P_{m}) - \alpha \chi^{2}(V_{Im}, V_{d})-\beta \chi^{2} (V_{Pm}, V_{d}) \\
- other ~ conditions
\end{split}
\end{equation}

\noindent
where $H$ is the entropy of a model map of the source, $\chi^{2}$ is a measure of the difference between the model and the observed visibilities (there are two $\chi^{2}$ terms, one for intensity and a second for polarisation), $\alpha$ and $\beta$ are the Lagrangian optimisation parameters and other conditions can also be included to represent additional constraints, such as the positivity of the intensity in the model map. A form of entropy suitable for polarisation emission developed by Gull and Skilling \cite{gull-skilling} and used by Holdaway \cite{holdaway} and Sault \cite{sault} is

\begin{equation}
\begin{split}
H=-\sum_{k} I_{k}(log(\frac{I_{k}}{IB_{k}e})+\frac{1+m_{k}}{2}log(\frac{1+m_{k}}{2})\\
+\frac{1-m_{k}}{2}log(\frac{1-m_{k}}{2}))
\end{split}
\end{equation}

\noindent
where $IB_{k}$ is the flux at pixel $k$ of a bias map (normally chosen to be a flat map with a total flux equal to the flux estimated for the source) and $I_{k}$ and $m_{k}$ are the Stokes intensity ($I$) flux and fractional polarisation, respectively, of pixel $k$.\\

The Gull and Skilling entropy, $H$, is a form of Shannon entropy (often used to describe the information content of a dataset), which has been generalised to include information on the polarisation of the data. An examination of the form of $H$ gives an indication as to how it will react to different types of sources. The Gull and Skilling entropy of a source that is described well by the bias map is high. A source which has low fractional polarisation (i.e. unordered magnetic field) will also have high Gull and Skilling entropy. The Gull and Skilling entropy is maximum for an unpolarised source that is identical to the bias map and this is the source that MEM will produce in the absence of any data that forces it to make a more complicated model (the $\chi^{2}$ terms in Eqn. \ref{jeqn} force MEM to make a model that maximises the Gull and Skilling entropy, but also reproduces the data to within noise levels). \\

By iteratively maximising $J$ in Eqn. (\ref{jeqn}), the MEM method develops a model of the source which maximises the Gull and Skilling entropy of the model (the model has lowest possible polarisation, and looks as much like the bias map as the data allows), while also reproducing the observed data to within noise levels. This results in a balance between entropy (representing the effects of unsampled visibilities and thermal noise) and fidelity to the observed data. This method of deconvolution, while not as straightforward as the CLEAN algorithm, is statistically and mathematically well-founded and can produce extremely well deconvolved maps comparable to, and in some cases better than, the CLEAN algorithm.\\

Unlike the standard CLEAN algorithm, MEM does not model the source as a series of delta functions. Instead MEM models the source as a continuous distribution - a more physically realistic model, but one which is computationally much more demanding. This increases the effective resolution of MEM, as it is not necessary to convolve the MEM model with the CLEAN beam. This means that the theoretical resolution of MEM is the Nyquist sampling theorem limit for the observation, although thermal and systematic noise may prevent drawing useful information at such small scales. It proves useful to convolve the MEM model map with a small beam to smoothen out these variations, although this limits the resolution of the resulting map. From experience, a beam of about $\frac{1}{2}$ to $\frac{1}{4}$ of the CLEAN beam works well for most sources.\\

MEM is also known for its mathematical property of "super-resolution". Unlike the CLEAN algorithm, MEM's resolution can was directly derived from Eq. (\ref{jeqn}) (see \cite{holdaway}) to be

\begin{equation}
x_{min}=\frac{1}{4 ~ u_{max}}
\end{equation}

\noindent
where $x_{min}$ is the resolution in the x direction and $u_{max}$ is the maximum baseline in the u direction (the same relation exists between the y direction in image space and the V direction in visibility space). This resolution is a factor of 4 below the best-case resolution expected from the Nyquist sampling limit, and therefore details at this resolution scale do not directly reflect information which has been recorded by the array. However, as MEM's model of the source as a continuous distribution is quite realistic, MEM can model the source at resolution levels below what has been observed. This modelling is done by creating a structure that can reproduce the data at the observed resolution levels while also having maximum Gull and Skilling entropy. In this way, MEM produces a conservative model of the source at resolutions below the Nyquist limit.\\

\begin{figure}

\begin{center}

\subfloat[Markarian 501 at 8.4 GHz. CLEAN image.]{
	\includegraphics[width=\columnwidth]{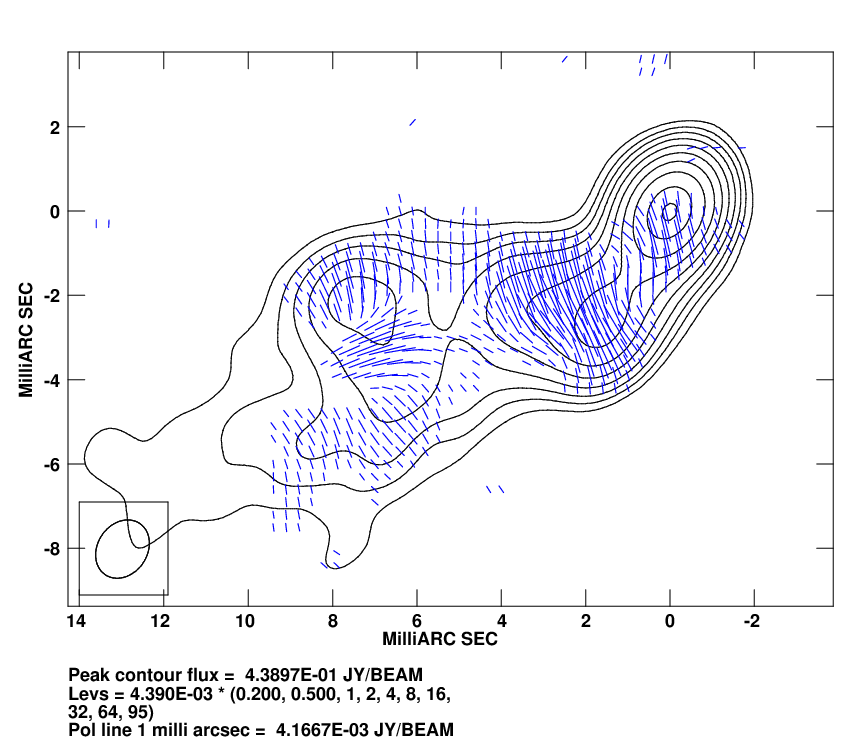}
	\label{markarian-fig-clean}
}

\subfloat[Markarian 501 at 8.4 GHz. MEM image.]{
	\includegraphics[width=\columnwidth]{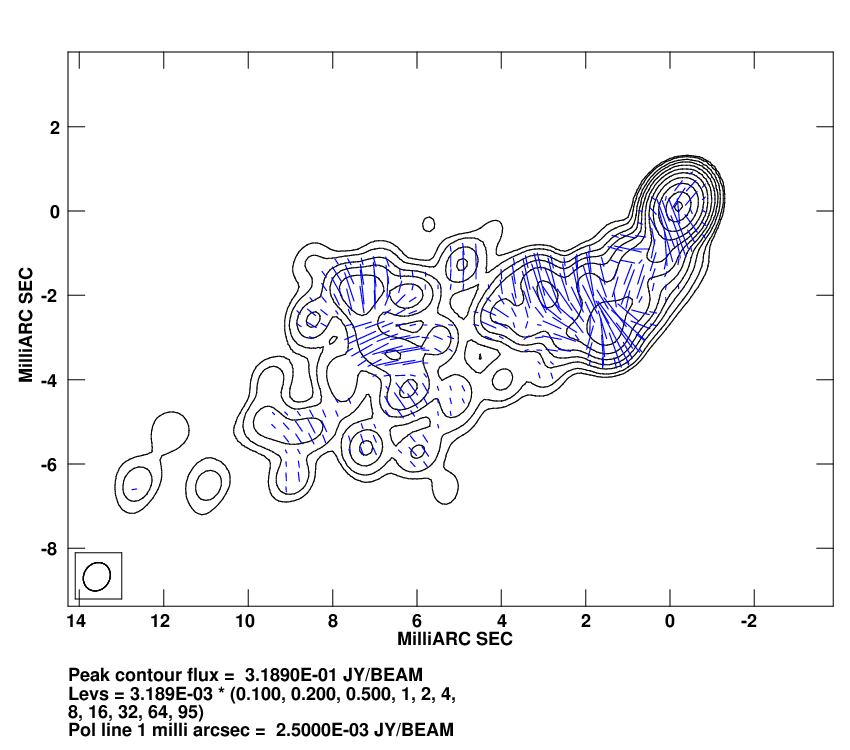}
	\label{markarian-fig-mem}
}

\caption{Markarian 501 at 8.4 GHz. The contours are Stokes I, the ticks indicate the direction of the observed polarisation. \textbf{	\ref{markarian-fig-clean}:} CLEAN image with a beam of 1.46 x 1.18 mas, $-33.85^{\circ}$ position angle. \textbf{\ref{markarian-fig-mem}:} MEM image, convolved with $\approx \frac{1}{2}$ of the CLEAN beam. For further information about the data used see ~\cite{pushkarev}.}
\label{markarian-fig}

\end{center}

\end{figure}

How well this "super-resolution" models detail at these lower levels is unknown, however the maps produced at these levels do not contain any obvious spurious features and appear to be a faithful extrapolation of the observed data. Monte Carlo simulations are being performed to test how well it performs on model sources (where the details of the source are known at sub-Nyquist resolutions).

\begin{figure*}

\subfloat[1633+382 CLEAN polarisation map]{
	\includegraphics[width=\columnwidth]{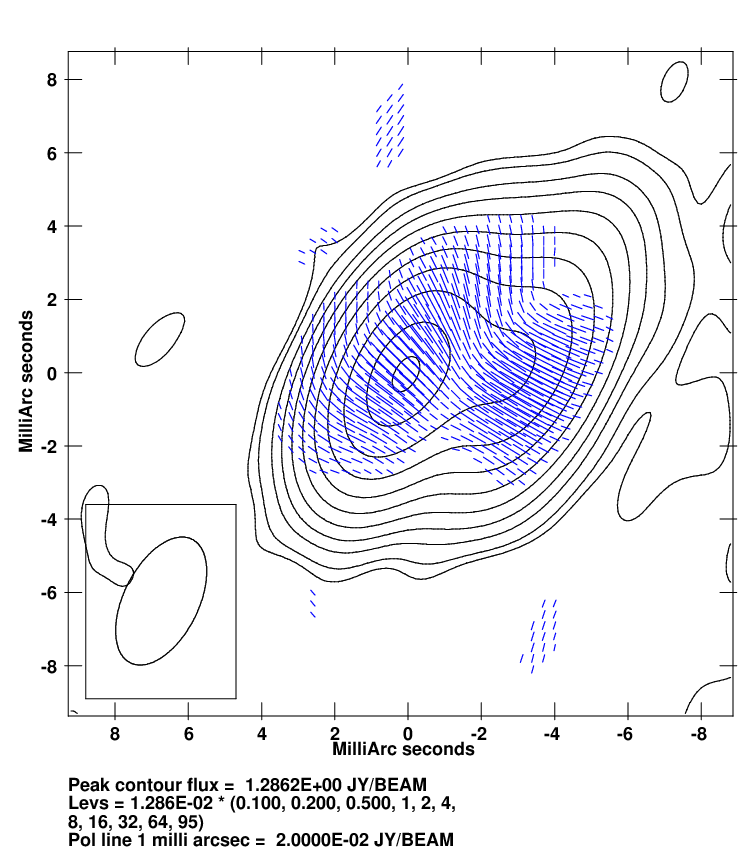}
	\label{1633+382-fig-pol-clean}
}
\hfill
\subfloat[1633+382 MEM polarisation map]{
	\includegraphics[width=\columnwidth]{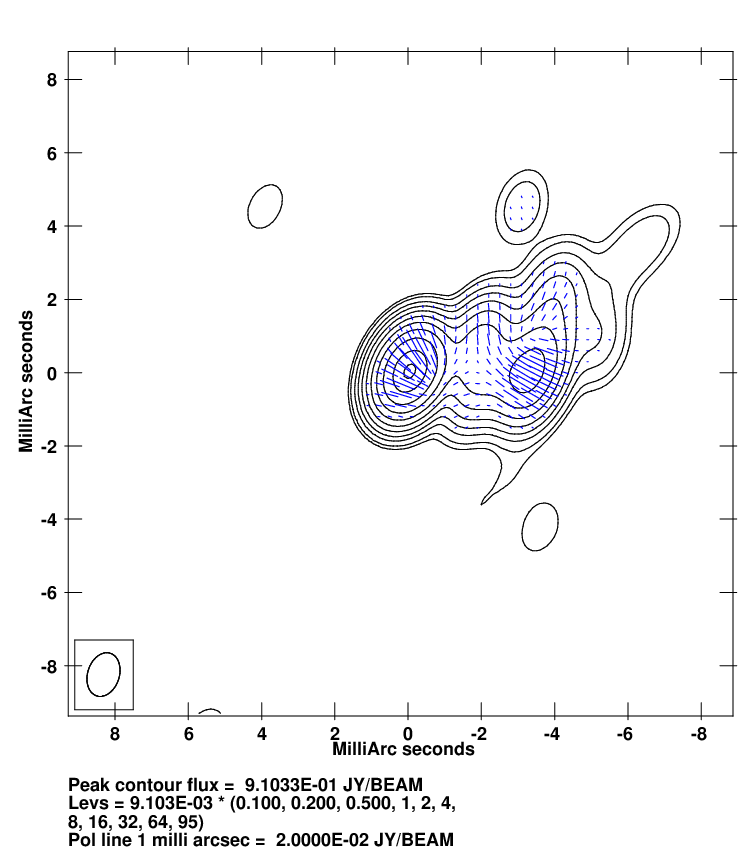}
	\label{1633+382-fig-pol-mem}
}

\subfloat[1633+382 CLEAN Faraday rotation measure map]{
	\includegraphics[width=\columnwidth]{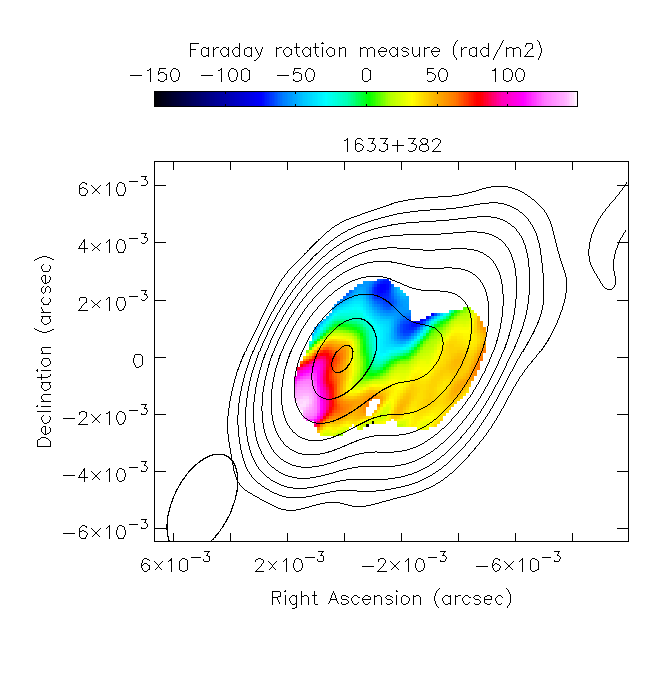}
	\label{1633+382-fig-rm-clean}
}
\hfill
\subfloat[1633+382 MEM Faraday rotation measure map]{
	\includegraphics[width=\columnwidth]{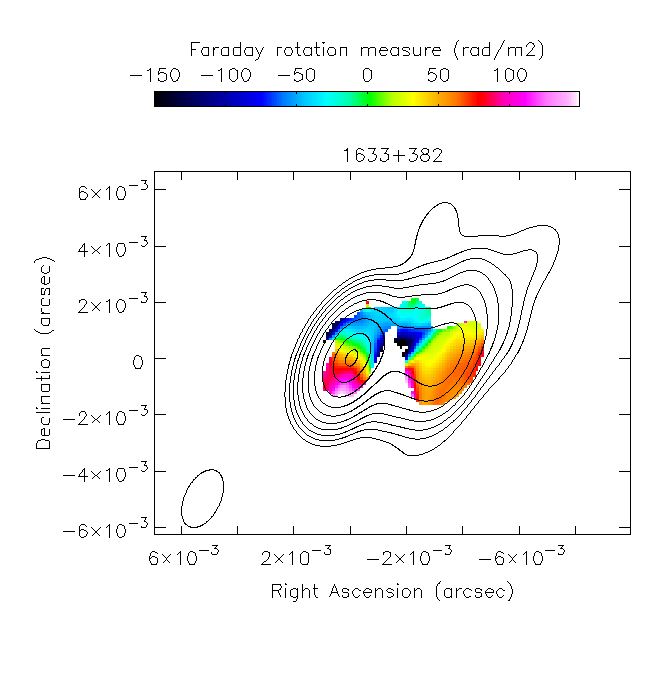}
	\label{1633+382-fig-rm-mem}
}

\caption{1633+382 at 4.6 GHz. The contours are Stokes I and the ticks indicate the direction of the observed polarisation. The colour scale is Faraday rotation measure in $rad/m^{2}$. The Faraday rotation measure maps were made with 6 frequencies running from 4.6 to 15.4 GHz. \textbf{\ref{1633+382-fig-pol-clean}:} CLEAN image with a beam of 3.77 x 2.04 mas, $-26.53^{\circ}$ position angle. \textbf{\ref{1633+382-fig-pol-mem}:} MEM image, convolved with $\approx \frac{1}{3}$ of the CLEAN beam. \textbf{\ref{1633+382-fig-rm-clean}:} CLEAN image with Faraday rotation, same beam. Note fewer contour lines are shown for clarity. \textbf{\ref{1633+382-fig-rm-mem}:} MEM image with Faraday rotation, convolved with $\approx \frac{1}{2}$ of the CLEAN beam. For further information about the data used see ~\cite{andrea}.}

\end{figure*}

\section{Implementation and Testing of New MEM Software}

A C++ program was written to implement a version of MEM for polarised data based on the MIRIAD task "PMOSMEM" \cite{miriad} and the AIPS task "VTESS" \cite{aips}, both of which use the Cornwell-Evans implementation of MEM \cite{cornwell-evans} but are unsuitable for deconvolving polarised VLBI data. VTESS does not support polarised data (although a AIPS related task, UTESS, implements a deconvolution method similar to MEM which can be used to deconvolve polarisation data), while PMOSMEM does support polarised data, but does not support VLBA data. The well-commented source codes of both tasks (PMOSMEM in particular) were extremely useful in writing a form of the algorithm which could handle polarised VLBA data. The software can be used by exporting  dirty maps for each Stokes parameter and the dirty beam for the observation as FITS files. The code takes in some basic parameters, such as an estimate of the total flux and the final (post-deconvolution) desired root-mean-squared noise. If the deconvolution has been successful, it produces FITS files which can then be imported into AIPS or CASA for viewing or further processing. The low level of human input required to run the software on a source makes it very suitable for use in a CASA or AIPS pipeline.\\

To ensure that the code was operating correctly and to characterise the behaviour of MEM based VLBI deconvolution, Monte Carlo simulations of the deconvolution of simulated sources are being performed. The code performs well deconvolving small Gaussians with realistic thermal noise, being able to recover the correct FWHM of the model Gaussian map. Further testing of the new polarisation MEM code using multiple Gaussian sources is currently underway.

\section{Results}

\subsection{Markarian 501}

Markarian 501 has an extended, bent jet with a "spine-sheath" polarisation structure visible in some places (see Fig. \ref{markarian-fig-clean} and \cite{pushkarev}), which was fitted using a helical magnetic-field model in \cite{murphy}. The MEM map shows this spine-sheath polarisation structure about 8 mas from the core more clearly (Fig. \ref{markarian-fig-mem}). The improved resolution of the MEM map separates the extended region of transverse polarisation (longitudinal magnetic field) about 4-mas from the core into several distinct regions with different polarisation orientations. The fan-like structure of the polarisation at the Southern edge of the jet in this region is suggestive of a longitudinal field component induced by local bending of the jet. The orthogonal orientation of the magnetic field in the inner jet is much more obvious in the MEM image.

\subsection{1633+382}

This jet also shows transverse polarisation structure, but the available CLEAN resolution is not sufficient to discern its nature (Fig. \ref{1633+382-fig-pol-clean}). The higher resolution offered by MEM shows the polarisation structure more clearly, in particular, the presence of orthogonal polarisation (longitudinal B field) along the top half of the jet and longitudinal polarisation (orthogonal B field) along the bottom half of the jet 1.5-2 mas from the core (Fig. \ref{1633+382-fig-pol-mem}).\\

The CLEAN Faraday rotation measure (RM) map constructed using VLBA data at 2-6cm (see Fig. \ref{1633+382-fig-rm-clean}) provided evidence for a transverse RM gradient across the jet, possibly due to a helical B field \cite{andrea}. The higher-resolution MEM RM map (Fig. \ref{1633+382-fig-rm-mem}) confirms this transverse RM structure, with a clear sign change in the RM from the Northern to the Southern part of the jet.\\

\section{Conclusions}

Software to implement a MEM-based deconvolution of VLBI polarimetry data has been written and is being tested with Monte Carlo simulations. Some first results using real, multi-wavelength VLBI polarisation data demonstrating its enhanced resolution over CLEAN have already been achieved. Future work will include multi-wavelength VLBI studies of a number of Active Galactic Nuclei at 2-6cm and 18-22cm. We intend to make our software available to the community once it has been fully tested and its operation well understood.

\section*{Acknowledgements}

This research has been funded by the Irish Research Council (IRC).

\end{document}